\documentclass[12pt,preprint]{aastex}

\usepackage{lscape}

\usepackage{graphicx}
\usepackage{CJK}

\shorttitle{Detection of methanol megamasers} \shortauthors{Xi Chen
et al.}

\begin{document}


\title {Detection of a methanol megamaser in a major-merger galaxy}
\author
 {Xi ~Chen\altaffilmark{1, 2}, Simon P. Ellingsen\altaffilmark{3}, Willem A. Baan\altaffilmark{1}, Hai-Hua Qiao\altaffilmark{1, 4},
Juan Li\altaffilmark{1, 2}, Tao An\altaffilmark{1, 2}, and Shari L.
Breen\altaffilmark{5}}

\altaffiltext{1} {Shanghai Astronomical Observatory, Chinese Academy
of Sciences, Shanghai 200030, China; chenxi@shao.ac.cn}

\altaffiltext{2} {Key Laboratory of Radio Astronomy, Chinese Academy
of Sciences, Nanjing, JiangSu 210008, China}

\altaffiltext{3} {School of Physical Sciences, University of
Tasmania, Hobart, Tasmania, Australia}

\altaffiltext{4} {University of Chinese Academy of Sciences, 19A
Yuquanlu, Beijing 100049, China}

\altaffiltext{5} {CSIRO Astronomy and Space Science, Australia
Telescope National Facility, PO Box 76, Epping, NSW 1710, Australia}

\label{firstpage}


\begin{abstract}

We have detected emission from both the $4_{-1}\rightarrow3_{0}$ E
(36.2~GHz) class I and $7_{-2}\rightarrow8_{-1}$ E (37.7~GHz) class
II methanol transitions towards the centre of the closest
ultra-luminous infrared galaxy Arp\,220. The emission in both the
methanol transitions show narrow spectral features and have
luminosities approximately 8 orders of magnitude stronger than that
observed from typical class~I methanol masers observed in Galactic
star formation regions.  The emission is also orders of magnitude
stronger than the expected intensity of thermal emission from these
transitions and based on these findings we suggest that the emission
from the two transitions are masers. These observations provides the
first detection of a methanol megamaser in the 36.2 and 37.7~GHz
transitions and represents only the second detection of a methanol
megamaser, following the recent report of an 84~GHz methanol
megamaser in NGC1068. We find the methanol megamasers are
significantly offset from the nuclear region and arise towards
regions where there is H$\alpha$ emission, suggesting that it is
associated with starburst activity. The high degree of correlation
between the spatial distribution of the 36.2~GHz methanol and X-ray
plume emission suggests that the production of strong extragalactic
class I methanol masers is related to galactic outflow driven shocks
and perhaps cosmic rays. In contrast to OH and H$_{2}$O megamasers
which originate close to the nucleus, methanol megamasers provide a
new probe of feedback (e.g. outflows) processes on larger-scales and
of star formation beyond the circumnuclear starburst regions of
active galaxies.

\end{abstract}


\section{Introduction}

The isotropic luminosity of molecular megamasers (MM) is more than a
million times greater than that of masers observed toward Galactic
star forming regions. MM emission from hydroxyl (OH) and
formaldehyde (H$_2$CO) are observed to trace the enhanced star
formation regions in the central molecular zones (CMZ) of luminous
infrared galaxies (LIRGs) (Baan 1989; Darling \& Giovanelli 2002;
Baan et al. 1993). In contrast, H$_2$O MM are exclusively associated
either with the circumnuclear accretion discs or jets of active
galaxies (Miyoshi et al. 1995) rather than mergers or LIRGs,
although kilomasers, which have luminosities of order a thousand
times greater than that of typical Galactic masers can be found in
LIRGs (e.g., Darling et al. 2008).

The intense maser emission observed from many transitions of
methanol (CH$_3$OH) within the Milky Way has been widely used to
study the formation and evolution of the most massive stars (e.g.
Green et al. 2009; Chen et al. 2014). Galactic methanol maser
transitions are empirically divided into two classes (Menten 1991).
Class~I methanol masers are produced in regions where molecular gas
is mildly shocked by processes such as outflows (e.g. Kurtz et al.
2004; Chen et al. 2009, 2011) or expanding H{\sc ii} regions
(Voronkov et al. 2010) . They are pumped through collisional
processes. Class~II methanol masers are found in the molecular gas
close to young high-mass stars and are pumped by far-infrared
radiation (Cragg et al. 2005).

Attempts to detect methanol masers in galaxies beyond the Milky Way
have had limited success. Extragalactic methanol masers have only
been detected in four galaxies, three of these being nearby galaxies
-- the Large Magellanic Cloud (LMC) and M31 for the 6.7 and/or
12.2~GHz class~II transition detections (Green et al. 2008;
Ellingsen et al. 2010; Sjouwerman et al. 2010), and NGC 253 for the
36.2~GHz class~I transition detection (Ellingsen et al. 2014). The
isotropic luminosity from these detected transitions is at most tens
of thousands of times higher than observed for typical Galactic
masers. Hence they fall several orders of magnitude short of being a
megamaser. To date, more than one hundred galaxies exhibiting OH or
H$_{2}$O megamaser emission have been searched for methanol
megamaser emission from the class~II 6.7 GHz transition, but there
are no reported detections (Ellingsen et al. 1994; Phillips et al.
1998; Darling et al. 2003). However, recent detections of methanol
maser emission from the class~I 36.2~GHz transition towards the
centre of the Milky Way (Yusef-Zadeh et al. 2013) and starburst
galaxy NGC 253 (Ellingsen et al. 2014) strongly suggests that
methanol megamaser emission from this transition may be detectable
in the central region of galaxies. This is further supported by the
recent report of methanol megamaser emission from the
$5_{-1}\rightarrow4_{0}$ E (84.5~GHz) transition towards the Seyfert
galaxy NGC\,1068 (Wang et al. 2014).

In this paper we report the first detection of both class~I and
class II methanol megamaser emission from the 36.2 and 37.7~GHz
transitions towards the central region of Arp\,220, the closest
ultra-luminous infrared galaxy (adopted distance 77.6 Mpc; 1$''$ =
376 pc), and an advanced merger system.

\section{Observations}

\begin{figure}[!th]
\begin{center}
\scalebox{0.5}[0.5]{\includegraphics[100,20][500,420]{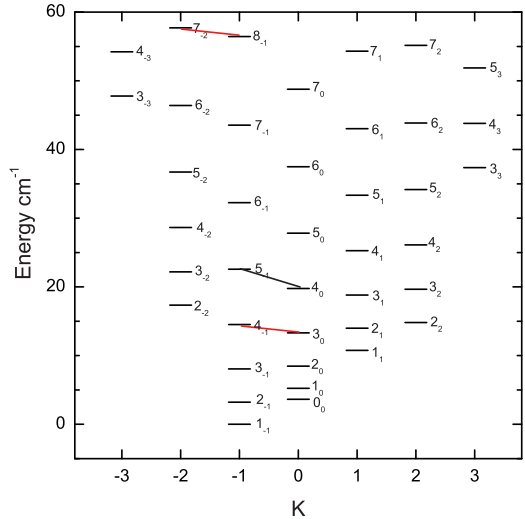}}
  \caption{The rotational energy levels of E-type methanol produced from CDMS catalog (M\"{u}ller et al.
2001). The transitions observed in this study are marked in red, the
transition detected in NGC 1068 by Wang et al. (2014) is marked in
black.} \label{fig:energy}
\end{center}
\end{figure}

The observations were made using the ATCA in the H168 array
configuration (baseline lengths between 61 and 192 m) on 2014 March
26. The synthesised beam width for the observations at 36 GHz was
approximately $\sim9''\times6.6''$. The data were collected using
the Compact Array Broadband Backend (CABB; Wilson et al. 2011)
configured with $2\times2.048$~GHz bands, centered on frequencies of
35.3 and 37.3~GHz respectively.  The two bands cover the rest
frequencies of the $4_{-1}\rightarrow3_{0}$ E class I and
$7_{-2}\rightarrow8_{-1}$ E class II transitions of methanol which
are at frequencies of 36.169265 and 37.703700~GHz, respectively
(M\"{u}ller et al. 2004). Figure~\ref{fig:energy} shows a rotational
energy level diagram of E-type methanol with the targeted
transitions marked. Each of the CABB bands has 2048 spectral
channels each of 1 MHz bandwidth, corresponding to a velocity width
of 8.3 km s$^{-1}$. The observing strategy involved three cycles of
10 minutes onsource on Arp 220 (pointing centre $\alpha=$15:34:57.2
$\beta=$+23:30:12 (J2000)) interleaved with 2 minute observations of
a phase calibrator (1548+056). Thus a total onsource time of 30
minutes on Arp 220 were made. PKS\,B1938-634 and PKS\,B1921-293 were
observed as the amplitude and bandpass calibrators, respectively.


The data reduction was performed with {\sc MIRIAD} using the
standard techniques for ATCA observations. Atmospheric opacity
correction was undertaken and the accuracy of the absolute flux
density calibration is estimated to be $\sim$30\%. Continuum
emission from the central region of Arp\,220 was used to produce a
model for a number of iterations of phase-only self-calibration. The
data for the 36.2 and 37.7~GHz methanol transitions was extracted
from the self-calibrated dataset through continuum subtraction and
imaged at a spectral resolution of 20 km s$^{-1}$. The resulting RMS
noise in a single spectral channel for the 36.2 and 37.7~GHz
methanol transitions was $\sim$ 4 mJy beam$^{-1}$.

In addition, observations which adopted the same CABB configuration
setup and observing strategy as used in the March observations were
also made in a Directors' time allocation on 2014 May 16 with the
1.5D array configuration (baseline lengths between 107 and 1469 m;
corresponding to a synthesised beam width of $\sim4''\times0.5''$).
A further 70 minutes on-source on Arp 220 were obtained.

\section{Results}

Continuum emission was detected towards the center of Arp\,220 as a
point source with an integrated intensity of 33.3 mJy at H168 array
observations. This measured intensity is similar to that ($\sim$ 60
mJy) recently measured at 32 GHz with the JVLA (Barcos-Mu\~{n}oz et
al. 2014) within a factor of $\sim$ 2. Using the data collected from
this array, we imaged the barycentric velocity range 4600--6200 km
s$^{-1}$ for both the 36.2 and 37.7~GHz methanol transitions with a
velocity resolution of 20 km s$^{-1}$. Other molecular species
within the central region of Arp\,220 (e.g., CO and HCN) show
emission over a velocity range from $\sim$ 5000 -- 5900 km s$^{-1}$
(e.g., Mart\'{\i}n et al. 2011; Greve et al. 2009). We summed the
spectral channels  within this velocity range to construct the
integrated emission in the two methanol transitions. Figure 2 shows
both the 36~GHz continuum emission and integrated methanol emission
of two methanol transitions with the optical H$\alpha$+[N {\sc ii}]
$\lambda\lambda$6548, 6583 intensity map (Taniguchi et al. 2012) as
the background. The optical H$\alpha$ emission is attributed to the
ongoing starburst activity. From Figure 2, it can be seen that the
majority of the emission from both the methanol transitions are
consistent with the distribution of the optical H$\alpha$ emission,
suggesting that the advanced merger provides the physical conditions
required to excite maser emission for the two methanol transitions.
Notably, strong H$\alpha$ emission is present in the two starburst
nuclei of Arp 220, however no significant methanol emission was
detected towards the central region of Arp\,220 wherein more active
starburst activity occurs. We discuss possible reasons for this in
Section 4.2.

At the angular resolution ($\sim7''$) of the ATCA H168 array
observations, two significant regions of 36.2~GHz methanol emission
were detected. They lie southeast and northwest of the Arp\,220
central region. The total extent of the methanol emission region is
approximately 20\arcsec\ which corresponds to 7.5 kpc for an assumed
distance of 77.6 Mpc. Three significant regions of 37.7~GHz methanol
emission were detected with offsets from the center of Arp\,220
ranging from 7--15 \arcsec (corresponding to 3--6 kpc).


\begin{figure*}[!t]
\begin{center}
\scalebox{2}[2]{\includegraphics[30,0][200,110]{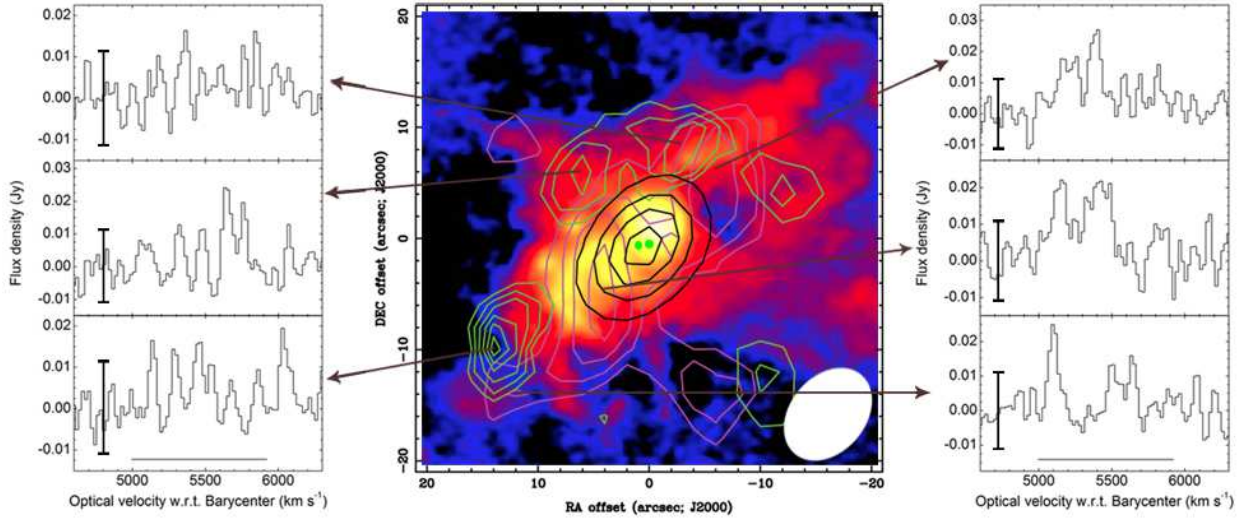}}
  \caption{The continuum and methanol emission in Arp\,220 taken in H168 array observations.
  The central panel displays the ATCA images of the 36~GHz continuum
emission (black contours), and the integrated methanol emission from
the 36.2~GHz (magenta contours) and the 37.7~GHz (green contours)
transitions. The background image is the continuum-subtracted,
smoothed optical H$\alpha$+[N {\sc ii}] $\lambda\lambda$6548, 6583
intensity image with color ranges from $1\times10^{-19}$ to
$5\times10^{-16}$ erg cm$^{-2}$ s$^{-1}$ (0.20 arcsec)$^{-2}$
(Taniguchi et al. 2012) . The synthesized beam of the ATCA
observations ( $\sim$9.0$''$ $\times$ 6.6$''$ with a P.A. of
-38$^\circ$ at 36~GHz) is shown in the bottom-right corner. The
contour levels for both integrated methanol transitions start at
2.5$\sigma$ and have increments of 1$\sigma$ ($1\sigma=$0.4 Jy
beam$^{-1}$ km s$^{-1}$). The contour levels of the continuum map
are at 20, 40, 60 and 80 $\sigma$ ($1\sigma=$0.35 mJy beam$^{-1}$).
The two green points near the center of the image mark the positions
of the two nuclei (separated by $\sim1''$) of Arp 220 (Sakamoto et
al. 2009). The left and right panels display the integrated spectra
of the 37.2~GHz class~II and 36.2~GHz class~I methanol emissions (at
a velocity resolution of 20 km s$^{-1}$) towards each of the
dominant emission region, which was formed from the image cube by
integrated over a rectangular region in right ascension and
declination area enclosing the significant emission in the
integrated intensity image. The $3\sigma$ significance level of each
spectrum is given with vertical solid line. The horizontal solid
line in the bottom of the two panels represents the observed
velocity range the thermal emission lines (e.g. CO and HCN; see
Greve et al. 2009) observed in Arp 220.}
\end{center}
\end{figure*}

We extracted the integrated spectrum of the two methanol transitions
towards the regions where significant emission is observed in the
integrated intensity images, and show them in the left and right
panels of Figure 2. The integrated spectra for these transitions
exhibit spectral features with linewidths of several tens of km
s$^{-1}$ towards each of these significant methanol emission
regions. The integrated spectrum over all these emission regions for
the 36.2~GHz transition (presented in Figure 3) shows two main broad
(110--180 km s$^{-1}$) spectral components. The spectrum of the
37.7~GHz transition integrated over all emission regions (Fig. 3)
shows six narrow (30--60 km s$^{-1}$) features covering the velocity
range  of 5000 -- 5900 km s$^{-1}$ (Greve et al. 2009). An
additional redshifted feature for the 37.7~GHz transition at a
velocity of $\sim$ 6000 km s$^{-1}$ may correspond to a separate
outflow component. We performed Gaussian fitting of these spectral
components and the parameters of the fits obtained are listed in
Table 1.


\tabletypesize{\scriptsize} \setlength{\tabcolsep}{0.05in}
\begin{deluxetable}{lccccccc}
\tablewidth{0pt} \tablecaption{Parameters of Gaussian fits to
methanol spectral features detected at H168 observations.}
\tablehead{ & \multicolumn{4}{c}{Gaussian Fits} \\
\cline{2-5}
 Transition  & Velocity &  FWHM width  & Integrated flux & Peak flux density   \\
 &  (km s$^{-1}$) & (km s$^{-1}$) &  (Jy km s$^{-1}$) & mJy}
\startdata
 36.2~GHz     & 5155(14) & 110(28) & 5200(1100) & 44 \\
              & 5411(18) & 188(42) & 7500(1600) & 37 \\
\cline{1-5}

 37.7~GHz      & 5138(13) &  65(14) & 1.7(7.1)  & 24 \\
               & 5346(7)  &  53(15) & 2.4(6.3)  & 42 \\
               & 5487(7)  &  48(18) & 1.9(6.1)  & 37 \\
               & 5675(11) &  90(28) & 3.6(9.3)  & 37 \\
               & 5773(7)  &  36(16) & 1.4(6.0)  & 36 \\
               & 5861(5)  &  35(15) & 1.7(5.4)  & 46 \\
               & 6055(13) & 105(30) & 3.5(9.3)  & 31 \\

\enddata

\end{deluxetable}

For comparison, the spectra of the two transitions measured from the
higher angular ($\sim 1''$) and spectral resolution observations
taken in the 1.5D array configuration (see Section 2) are also shown
in Figure 3. Their spectra at a velocity resolution of 10 km
s$^{-1}$ were extracted from the uv-data by vector averaging towards
the regions where methanol emission was observed in the images
obtained from the H168 array observations. The higher angular
resolution observations show the presence of narrower spectral
components with line widths of less than about 10 km s$^{-1}$ for
the both methanol transitions as they were only detected in $1-2$
channels. The majority of the methanol emission detected for the two
transitions in the H168 array observations were not detected in the
1.5D array data. However, because of the limited uv-coverage of the
1.5D array data we can not obtain useful images and can not
determine if the non-detection is because the emission is resolved
by the longer baseline, or if it is due to atmospheric
decorrelation.


\begin{figure}[!t]
\begin{center}
\scalebox{0.45}[0.45]{\includegraphics[100,00][500,450]{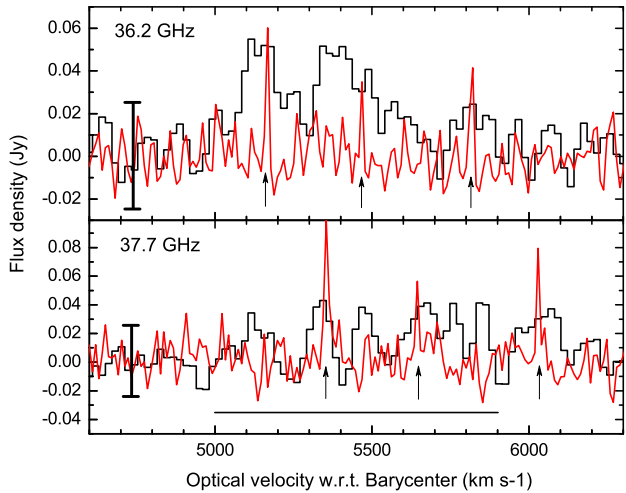}}
 \caption{Integrated spectra of the 36.2 and 37.7~GHz methanol
transitions. For each transition the black solid line represents the
total integrated flux observed in the H168 array for a spectral
resolution of 20 km s$^{-1}$ summed over the significant methanol
emission regions (shown in Fig. 2). The red solid line represents
the total integrated flux of the same regions observed at higher
spatial resolution (with the ATCA in the 1.5D array) and for a
spectral resolution of 10 km s$^{-1}$. The upward arrows mark the
velocities of the narrow spectral features in each of the plots. The
spectra from the H168 and 1.5D array observations have a similar rms
noise and the same $3\sigma$ significance level of spectra is given
for them with vertical solid line in each panel.}
\end{center}
\end{figure}

\section{Discussion}

\subsection{The nature of the 36.2 and 37.7~GHz methanol emission}

Within the Galaxy the 36.2 and 37.7~GHz masers are observed towards
star formation regions (see Voronkov et al. 2014; Ellingsen et al.
2011, 2013). Based on the luminosity, narrow line widths ($<$ 10 km
s$^{-1}$) and significant deviation from thermal intensities,
Ellingsen et al. (2014) suggest that the 36.2~GHz methanol emission
in the central region of NGC\,253 are masers. The spectrum of the
36.2~GHz methanol emission detected towards Arp\,220 shows much
broader spectral features than those in NGC 253, however, they are
still 2-3 times narrower than the other thermal molecules detected
in Arp\,220 which have typical linewidth $>300$ km s$^{-1}$ (e.g.
Greve et al. 2009). The 37.7~GHz methanol emission in Arp\,220 shows
narrower spectral features than the 36.2~GHz transition, with
typical velocity widths of $\sim$ 50 km s$^{-1}$. The presence of
very narrow spectral features (with line width of a few km s$^{-1}$)
in the two transitions at angular resolutions of around $1''$ (taken
in 1.5D array) provides strong evidence that at least some of the
detected emission is from masers.

Adopting the typical excitation conditions (column density
$7\times10^{15}$ cm$^{-2}$ and excitation temperature 17 K) observed
for thermal methanol emission in the central region of Arp\, 220
(Mart\'{\i}n et al. 2011), the predicted integrated intensity for
the 36.2 and 37.7 GHz transitions are 8.2 and 0.2 K km s$^{-1}$,
respectively. Whereas the observed integrated intensity of the
methanol emission from these two transitions from the ATCA
observations (at $\sim$7$''$ angular resolution) is 280 and 360 K km
s$^{-1}$, respectively, approximately 34 and 1800 times larger than
expected if the emission is thermal. This provides further evidence
that the emission from these two methanol transitions is produced by
masers.

The integrated intensity of the observed 36.2 and 37.7~GHz methanol
emission in Arp\,220 corresponds to an isotropic luminosity of
$\sim$2300 L$_{\odot}$ and $\sim$3000 L$_{\odot}$, respectively.
Compared with the isotropic luminosity of typical Galactic methanol
masers in the 36.2 and 37.7~GHz transitions of $\sim$10$^{-5}$
L$_{\odot}$, the emission lines are approximately eight orders of
magnitude stronger than typical Galactic class~I and class~II
methanol masers, hence they represent the first detection of
methanol megamasers in both the 36.2 and 37.7 GHz transitions.  Wang
et al. (2014) recently reported the detection of 84.5~GHz methanol
emission from NGC\,1068 with a luminosity of a few million times
that of typical Galactic class~I methanol masers. Therefore the
luminosity of 36.2 GHz methanol in Arp\,220 is two orders of
magnitude stronger than that of the 84.5 GHz methanol in NGC\,1068.
The 84.5 and 36.2~GHz methanol transitions are from the same
transition family (see Figure~\ref{fig:energy}) and in most Galactic
sources where both have been observed the 36.2~GHz emission is
significantly stronger.

\subsection{The pumping mechanism for the two type methanol masers}

Essentially all of the molecular emission (e.g. CO) detected in
previous observations of Arp\,220 originates from the CMZ region
which has an extent of about 1 kpc and encompasses the two merger
nuclei (Sakamoto et al. 2009). But the 36.2 and 37.7~GHz methanol
emission in Arp\,220 originates predominantly outside the CMZ, and
this is similar to the situation observed for the 36.2~GHz methanol
transition in NGC\,253. The mechanism proposed by Yusef-Zadeh et al.
(2013) to explain the 36.2~GHz methanol emission in the Milky Way
CMZ is that high cosmic ray intensities (compared to the Galactic
disk) play the key role in releasing methanol into the gas phase
from cold dust grains and enhance the abundance of methanol,
however, once the released into the gas-phase methanol will be
rapidly destroyed if the high cosmic ray intensity is too high.
Ellingsen et al. (2014) suggested that this mechanism is consistent
with the observed distribution of 36.2~GHz methanol emission in
NGC\,253, with the absence of any emission in the inner 100~pc being
due to extreme cosmic ray fluxes. However, at present there is no
direct observational evidence to support this speculation.

Figure 4 presents the methanol emission overlaid on the X-ray
emission image of McDowell et al. (2003). This figure clearly shows
that the 36.2~GHz methanol emission is highly correlated with the
diffuse X-ray emission in the north-western and east-southern
regions, referred to as the X-ray ``plume'' by McDowell et al.
(2003). The central nuclear region hosts higher intensity, more
compact X-ray emission, but there is no significant 36.2~GHz
methanol emission associated with that gas. We expect the X-ray
emission to reflect the cosmic ray intensity in Arp\,220. Hence, the
observed close correspondence between the 36.2~GHz methanol and
diffuse X-ray emission and the anti-correlation with high-energy
X-ray emission supports the hypothesis that cosmic ray intensity
governs the locations where there is a significant abundance of
gas-phase methanol. The X-ray plume in Arp\,220 is thought to be the
result of a starburst-generated superwind (McDowell et al. 2003).
The superwind represents regions with both rapidly outflowing and
inflowing shock-heated gas, which may provide the collisional
pumping agent for the widespread 36.2~GHz class~I methanol emission.
The methanol is pumped in the expanding shocks surrounding the X-ray
cocoon and the distributed emission results from amplification of
background or embedded radio continuum. The shocks within Galactic
star formation regions produce class~I maser emission distributed on
scales of around 1~pc, while the plume region in Arp\,220 has a
scale two to three orders of magnitude larger.


\begin{figure}[!t]
\begin{center}

 \scalebox{1.6}[1.6]{\includegraphics[20,5][100,125]{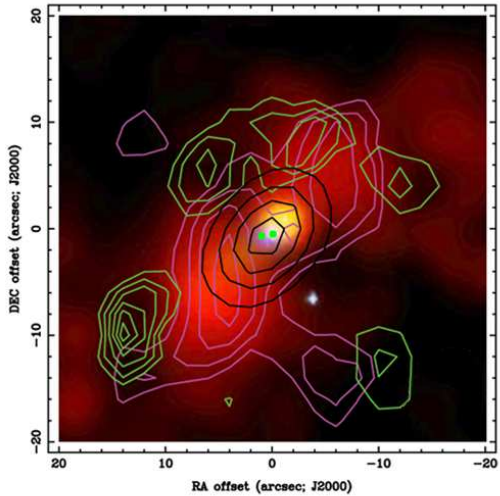}}

 \caption{Comparison of the methanol and X-ray emission. The contour
images obtained from the ATCA as presented in Fig. 2 superposed on
the background of the X-ray emission where red represents the
0.2--1.0 keV band, green the 1.0--2.0 keV band, and blue the
2.0--10.0 keV band (McDowell et al. 2003).}

\end{center}
\end{figure}

Within the Milky Way class~II methanol masers and OH masers are
often co-located on arcsecond scales, although they are not
cospatial on milliarcsecond scales (Menten et al. 1992; Li et al.
2012), and both species are pumped by emission at far-infrared (FIR)
wavelengths produced by warm dust (Cragg et al. 2005). In Arp\,220,
the OH and H$_{2}$CO megamaser emission is observed towards the
positions of the two nuclei (see Figs. 2 and 4), and are thought to
result from FIR pumping and amplification of background radio
continuum (Rovilos et al. 2003; Baan 1985, 1989). There is one
previous detection of a class~II methanol transition towards
Arp\,220, with a detection of the 6.7~GHz transition in absorption
(rather than as a maser) (Salter et al. 2008). The 37.7~GHz class~II
methanol emission regions observed in Arp\,220 do not spatially
correlate with the X-ray structure, but rather surround/straddle the
region. As for the 36.2~GHz transition, 37.7~GHz emission arises on
much larger scales than the OH and H$_{2}$CO masers and is
significantly offset from the nucleus. The 37.7~GHz emission regions
represents the location of cooler and denser gas at the edge of the
diffuse X-ray emission, possibly in regions of triggered star
formation, which may host gas with enhanced methanol abundance
either produced in-situ or driven there by outflows from the plume
region. However, existing radio and infrared observations of
Arp\,220 show no significant counterparts from these regions.

\section{Conclusion}

We have detected the first 36.2 and 37.7 GHz methanol megasers
towards the major merger galaxy Arp\,220. The characteristics of the
spectral lines and the spatial distribution of the 36.2 and 37.7~GHz
methanol megamaser emission is significantly different from that of
the megamasers of other species and the thermal molecular emission
in Arp\,220. The observed methanol megamasers trace larger-scale
feedback (e.g. outflows) and star formation in circumnuclear
starburst regions of a merger system, in contrast to other species
of megamaser which originate close to the nucleus. Therefore,
methanol megamasers will trace a different physical environment and
provide diagnostic and structural information that is complementary
to those of other molecular transitions observed in starbursts and
the feedback systems in active galaxies.

\section{Acknowledgements}


Our results are based on the observations made using the Australia
Telescope Compact Array (ATCA), which is part of the Australia
Telescope National Facility funded by the Commonwealth of Australia
for operation as a National Facility managed by CSIRO. We thank Dr.
Taniguchi Y. for his providing optical H$\alpha$ line data. XC
acknowledges supports from the National Natural Science Foundation
of China (11133008 and 11273043), the Strategic Priority Research
Program of the Chinese Academy of Sciences (CAS; Grant No.
XDA04060701), Key Laboratory for Radio Astronomy, CAS. WAB has been
supported as a Visiting Professor of the Chinese Academy of Sciences
(KJZD-EW-T01) Shari Breen is the recipient of an Australian Research
Council DECRA Fellowship (project No. DE130101270).

\end{document}